\documentclass{aa}  

\usepackage{graphicx}
\usepackage{txfonts}
\usepackage{natbib}
\bibpunct{(}{)}{;}{a}{}{,} % to follow the A&A style
\usepackage[version=3]{mhchem}
\usepackage[ampersand]{easylist}
\usepackage{amsmath, amssymb}
\usepackage{amsmath}

\begin{document}

   \title{Wavelength-dependent UV photodesorption of pure N$_2$ and O$_2$ ices}

   %\subtitle{I. Overviewing the $\kappa$-mechanism}

   \author{ E.C.~Fayolle\inst{1}
          \and
          M.~Bertin\inst{2}
          \and 
          C.~Romanzin\inst{3}
          \and
          H.A.M~Poderoso\inst{2}
          \and
          L.~Philippe\inst{2}
          \and
           X.~Michaut\inst{2}
           \and
           P. Jeseck\inst{2}
           \and
           H.~Linnartz\inst{1}
           \and
           K.I.~\"Oberg\inst{4}
           \and
           J.-H.~Fillion\inst{2}
          }

   \institute{Sackler Laboratory for Astrophysics, Leiden Observatory, Leiden University, P.O. Box 9513, 2300 RA Leiden, The Netherlands\\ \email{fayolle@strw.leidenuniv.nl}
    	\and
	Laboratoire de Physique Mol\'{e}culaire pour l'Atmosph\`{e}re et l'Astrophysique, UPMC Univ. Paris 6, CNRS-UMR7092,
              4 place jussieu, 75252 Paris, France
         \and
	    Laboratoire de Chimie Physique, UMR 8000 CNRS-Universit\'{e} Paris-Sud, 91405 Orsay, France
	\and
	   Departments of Chemistry and Astronomy, University of Virginia, Charlottesville, VA 22904, USA
             }

   \date{Received X; accepted Y}

% \abstract{}{}{}{}{} 
% 5 {} token are mandatory
 
  \abstract
   {Ultraviolet photodesorption of molecules from icy interstellar grains can explain observations of cold gas in regions where thermal desorption is negligible. This non-thermal desorption mechanism should be especially important where UV fluxes are high.}
   {N$_2$ and O$_2$ are expected to play key roles in astrochemical reaction networks, both in the solid state and in the gas phase. Measurements of the wavelength-dependent photodesorption rates of these two infrared-inactive molecules provide astronomical and physical-chemical insights into the conditions required for their photodesorption. }
   {Tunable radiation from the DESIRS beamline at the SOLEIL synchrotron in the astrophysically relevant 7 to 13.6 eV range is used to irradiate pure N$_2$ and O$_2$ thin ice films. Photodesorption of molecules  is monitored through quadrupole mass spectrometry. Absolute rates are calculated by using the well-calibrated CO photodesorption rates. Strategic N$_2$ and O$_2$ isotopolog mixtures are used to investigate the importance of dissociation upon irradiation.}
   {N$_2$ photodesorption mainly occurs through excitation of the b$\rm^1\Pi_u$ state and subsequent desorption of surface molecules. The observed vibronic structure in the N$_2$ photodesorption spectrum, together with the absence of N$_3$ formation, supports that the photodesorption mechanism of N$_2$ is similar to CO, i.e., an indirect DIET (Desorption Induced by Electronic Transition) process without dissociation of the desorbing molecule. In contrast, O$_2$ photodesorption in the 7~-~13.6~eV range occurs through dissociation and presents no vibrational structure.}
   {Photodesorption rates of N$_2$ and O$_2$ integrated over the far-UV field from various star-forming environments are lower than for CO. Rates vary between 10$^{-3}$ and 10$^{-2}$ photodesorbed molecules per incoming photon.}


   \maketitle
%
%________________________________________________________________

\section{Introduction}

In the cold and dense regions of the interstellar medium (ISM), molecules condense onto the surfaces of submicron-sized dust grains. Upon heating or non-thermal desorption, these molecules are released into the gas phase. Constraining these desorption mechanisms is crucial since they bridge solid state and gas phase chemistry in space. In regions where heating can be neglected, non-thermal desorption pathways are required to explain the presence of molecules in the gas phase that, without such mechanisms, should remain frozen. Non-thermal desorption has been proposed to take place in star-forming regions at every evolutionary stage to explain observations of cold gas: in prestellar cores \citep{2012ApJ...759L..37C}, in protostellar envelopes \citep{2010A&A...516A..57K} and in protoplanetary disks \citep[e.g.,][]{2000ApJ...544..903W,Dominik:2005bp,2011Sci...334..338H}. Non-thermal desorption in the ISM is induced by cosmic-rays, exothermic reactions, shocks, electrons, and photon irradiation. Experimental measurements of the efficiencies of these processes are a prerequisite to understand which of these pathways must be considered to predict molecular gas abundances.\

Precise measurements are especially important for abundant molecules that can play key roles in gas-grain chemical networks. Two of these species are N$_2$ and O$_2$ because of their role in the formation of N- and O-bearing molecules in the gas phase and in the solid state. For instance, the presence of N$_2$ in the solid state leads to the formation of NH$_3$ \citep[e.g.,][]{Aikawa:1997hk,Daranlot:2012cg}. In the gas phase, the N$_2$ derivative N$\rm_2H^+$ is a major tracer of dense cold cores \citep[e.g.,][]{Caselli:2002cb}. Molecular oxygen plays a role in water formation in the gas phase after dissociation into atomic oxygen but also in the solid phase \citep{Hollenbach:2008ho}. O$_2$ hydrogenation has been proposed as a starting point for water formation on grains by \cite{Tielens:1982tb} and proven experimentally to be effective by many studies \citep[e.g.,][]{Miyauchi200827,2008ApJ...686.1474I,Matar:2008cz}. The abundance of H$_2$O on the grains is thus intrinsically linked to the amount of O$_2$ ice, which is itself regulated by non-thermal desorption mechanisms at low temperatures.\

It is difficult to put observational constraints on molecular nitrogen and oxygen abundances and guide gas-grain modeling since these species do not possess permanent dipole moments. Some O$_2$ electronic transitions have been observed in the far-UV (FUV) using Odin \citep{2003A&A...402L..77P} and some tentative detections have been reported in the millimeter range \citep{2011ApJ...737...96G,Liseau:2012bj}, resulting in low gas-phase abundances in specific environments. Molecular nitrogen has been detected by \cite{Knauth:2004el} in the FUV but its abundance is mainly inferred via N$_2$H$^+$ detections in the millimeter range. Since observational data on such species are rare and limited to specific ISM regions, accurate laboratory data are necessary to support gas-grain codes, including desorption efficiencies.\

Among the non-thermal desorption mechanisms, UV photodesorption has been studied in detail for specific molecules by various groups. Photodesorption rates for CO \citep{Oberg:2007ip,Oberg:2009kca,2010A&A...522A.108M,Fayolle:2011eh,Bertin:2012wi}, H$_2$O \citep{1995Natur.373..405W,2009ApJ...699L..80Y,Oberg:2009kla}, CO$_2$ \citep{Oberg:2009kca,2012ApJ...761...36B}, and N$_2$ \citep{Oberg:2009kca} are available for gas grain codes \citep[for e.g.][]{2000ApJ...544..903W,Walsh:2010fl,2013ApJ...762...86V}. However, most of the previous studies provide photodesorption rates for photons at a specific energy, mainly Lyman-alpha (121.6 nm, 10.2 eV) or for broad-band Vacuum-UV (VUV) and details on the underlying mechanisms are limited. As shown in \cite{Fayolle:2011eh} and \cite{Bertin:2012wi}, studying photodesorption with respect to the incoming photon energy provides 1) photodesorption rates that can be used to predict the process efficiency in different phases of the ISM, taking the specific UV-field profile into account, and 2) detailed insights into the underlying molecular mechanism. Both kinds of information are needed to extrapolate photodesorption rates to astrophysical UV fluxes and time scales. In the case of CO ice, photodesorption occurs through an indirect DIET (Desorption Induced by Electronic Transition) process where surface CO desorbs through electronic excitation of subsurface molecules. This mechanism is only valid below 10 eV where photons do not induce CO dissociation. For other species (e.g. H$_2$O, CO$_2$, CH$_3$OH...), the dissociation of the parent molecules into radicals strongly affects the photodesorption mechanism \citep[e.g.,][]{1995Natur.373..405W,Oberg:2009kla}.\

 In the present study, N$_2$ and O$_2$ ice photodesorption is explored by performing wavelength-dependent irradiation experiments between 7 and 13.6 eV. This specific spectral window has been explored since it corresponds, for the upper limit, to the Lyman edge above which most of the radiation is absorbed by hydrogen and for the lower limit, to the minimum energy required for electronic excitation of most simple interstellar molecules. The experimental techniques that are used to measure the photodesorption rates of these non-IR active species are presented in the Method section. The results are presented separately for N$_2$ and O$_2$ in section \ref{Sec_res}. A discussion of the involved mechanisms, photodesorption rates, and recommendations for use in gas-grain codes is presented in section \ref{Sec_dis}, followed by the conclusions.

%__________________________________________________________________

\section{Method}

\subsection{Experiments}
	Experiments are performed on ices grown in the ultra-high vacuum set-up SPICES (detailed description available in \citealt{Bertin:2011ed}), which can reach low pressures down to 10$^{-10}$ mbar. Ices of $^{14}$N$_2$, $^{16}$O$_2$ (both from Air liquide, alphagaz 2, polluants $<$~1~ppm), $^{15}$N$_2$ (Eurisotop, 97\% $^{15}$N), and $^{18}$O$_2$ (Eurisotop, 97\% $^{18}$O) are grown diffusively with monolayer (ML) precision on either highly oriented pyrolitic graphite (HOPG) or a polycrystalline gold surface. Both substrates are mounted on the same rotatable cold head, which can be cooled down to temperatures as low as 14~K by a closed cycle helium cryostat. The set-up is equipped with a quadrupole mass spectrometer (QMS model 200 from Balzers) allowing for gas phase molecule detection via electron impact at 90~eV, as well as a Fourier-transform infrared spectrometer (Vector 22 model from Bruker) for reflection-absorption infrared spectroscopy (RAIRS) to probe changes in the condensed phase. Ice thicknesses are chosen to be between 30 and 60~ML, which is expected to be thicker than the photodesorption-active layers. The targeted ice thickness is achieved (within few tenths of an ML) by calibrating the rate and deposition time using the temperature program desorption (TPD) technique. The TPD curves are sensitive to the break between multilayers and monolayer desorption due to the difference in binding energy between ice-ice and ice-substrate. Isotopolog mixtures are produced in situ by using two deposition valves connected to the deposition tube. The desired mixing ratio is achieved by monitoring the partial pressure of the gases during deposition by mass spectrometry.\
	
	Photodesorption of ices is induced by UV photons from the undulator-based DESIRS beamline \citep{Nahon:2012dn} at the SOLEIL synchrotron. In the explored 7~--~13.6~eV range, the undulator provides continuous monochromatic light with an FWHM of about 1 eV, as well as higher harmonics that are suppressed when the beam traverses a specific gas filter. For an irradiation experiment at fixed energy requiring a high flux, specific energies are chosen using the undulator, and the photons are used straight from the gas filter resulting in wavelength-fixed fluxes between 10$^{13}$ and 5~$\times$~10$^{14}$~photons~s$^{-1}$, and measured by calibrated photodiodes before each irradiation experiment. To acquire wavelength-dependent photodesorption spectra, a higher energy resolution is obtained using the 6.65~m normal incidence monochromator
implemented on the beamline. It is possible to perform energy scans between 7 and 13.6~eV with a narrow bandwidth of typically 40~meV and intensities at the sample between 3 and 11.5~$\times$~10$^{12}$~photons~s$^{-1}$~cm$^{-2}$.\

	The photodesorption experiments are performed by hooking up the SPICES chamber directly to the beamline, thus avoiding any radiation cut-off due to optical components. The photon angle of incidence on the substrate is 45$^{\rm o}$, the UV beam spot is 0.7 $\pm$ 0.1 cm$^2$, and the light is in \emph{s}-polarized mode (perpendicular to the plane of incidence). The photodesorption efficiency is assessed over 7~--~13.6 eV by scanning the photon energy and simultaneously recording the relative level of photodesorbed molecules by mass spectrometry. Additional high flux experiments can be performed at fixed energy, and desorption is monitored simultaneously by RAIRS and QMS.\

	 Photodesorption data obtained from  N$_2$ or O$_2$ ices are found to be identical for films grown on HOPG and Au samples, demonstrating a negligible effect of the nature of the substrate. The photodesorption spectra recorded here are mainly obtained from ices deposited on HOPG, whereas the gold substrate has been specifically used to probe the irradiated ices by IR spectroscopy.

\subsection{Data analysis}
	
	In the case of the previous wavelength-dependent CO photodesorption studies \citep{Fayolle:2011eh,Bertin:2012wi}, the absolute photodesorption spectrum is found by quantifying the absolute
ice loss during irradiation at a single energy and high photon flux using infrared spectroscopy. This photodesorption value at single energy is used to scale the relative photodesorption spectrum obtained by mass spectrometry following the signal of the molecular ion CO$^+$ at low flux. This technique cannot, however, be directly used for N$_2$ nor O$_2$ since neither of them have permanent dipole moments and are thus not detectable in the infrared.\
	
	To derive absolute N$_2$ and O$_2$ photodesorption rates, the relative photodesorption spectra obtained from the QMS signals of the corresponding molecular ions (m/z=28 and 32, respectively) are scaled using the QMS to absolute photodesorption rate factor derived for CO and correcting it for the gas-specific ionization cross-sections\footnote{The ionization cross sections by electron impact at 90 eV used for calibration here are: 1.73~\AA$^2$ for CO and 1.78~\AA$^2$ for N$_2$ from \cite{Freund:1990hx} and 1.64~\AA$^2$ for O$_2$ from \citep{Straub:1996fp}} \citep{Freund:1990hx,Straub:1996fp} as 
		
	$$\rm Y_i = f_{i^+} \times I_{i^+} \;\; with \;\; f_{i^+} = f_{CO^+} \times \frac{\sigma_{i^+}}{\sigma_{CO^+}}$$
	
\noindent where Y$_{\rm i}$ is the absolute photodesorption rate of molecule i, I$_{\rm i^+}$ is the QMS signal of the corresponding molecular ion, f$_{\rm i^+}$ the QMS count to absolute photodesorption factor for species i, f$_{\rm CO^+}$ the same factor for CO derived by infrared spectroscopy during the same experimental runs, using identical QMS and light setting and $\sigma_{\rm i^+}$ and $\sigma_{\rm CO^+}$ are the ionization by electron impact cross-sections at 90 eV for species i and CO.\
	
	Using CO measurements to calibrate N$_2$ and O$_2$ photodesorption is a reasonable approach because N$_2^+$, O$_2^+$, and CO$^+$ fragments have very close masses. The apparatus function of the QMS (mainly dominated by the gain due to the secondary electron multiplier (SEM) and fragment transmission efficiency) should thus be similar (within few percents) for a given ionization energy. The gas phase velocity distribution of the desorbing species may influence the QMS detection efficiency. In the present experiments, velocities of the photodesorbed molecules were not specifically determined, but are expected to be of the same order of magnitude for CO, O$_2$, and N$_2$ that all three have very similar masses and ice binding energies. Consequently, the mass signal ratios of N$_2^+$ or O$_2^+$ over CO$^+$ only depends on the respective gas partial pressures and on the 90 eV ionization cross-sections by electron impact of the corresponding gases \citep{Kaiser:1995fl}.\

	Great care has been taken to insure that the QMS signal used to probe the photodesorption rates unambiguously originates in the parent desorbing molecules (i.e. N$_2$, O$_2$, or CO). The absence of QMS signals that could be attributed to N$_3^+$, O$_3^+$, or CO$_2^+$ in low light flux conditions experiments clearly indicates that the molecular ion signals are not contaminated by secondary fragmentation in the QMS of N$_3$, O$_3$, or CO$_2$, respectively ; the electron impact fragmentation patterns of these neutral species at 90 eV have been considered for this purpose \citep[for example, see][for detailed analysis of O$_3$ fragmentation]{Cosby:1993dw,Bennett:2005gn,2011PCCP...13.9469E}.

A first source of uncertainty in the photodesorption yields comes from the QMS signal-to-noise ratio and small residual gas fluctuations. The root mean square of the spectra reaches $\rm 4 \times 10^{-4}$ molecules~photon$^{-1}$ for $^{15}$N$_2$ and $\rm 6 \times 10^{-4}$ molecules~photon$^{-1}$ for O$_2$. In addition, there is a 40~\% systematic uncertainty inherent to the calibration method from relative QMS count to absolute photodesorption rates for CO \citep[see][]{Fayolle:2011eh}.

\section{Results} \label{Sec_res}

	\subsection{Nitrogen photodesorption}

		\begin{figure}
			\centering
 			\includegraphics[width=1\linewidth]{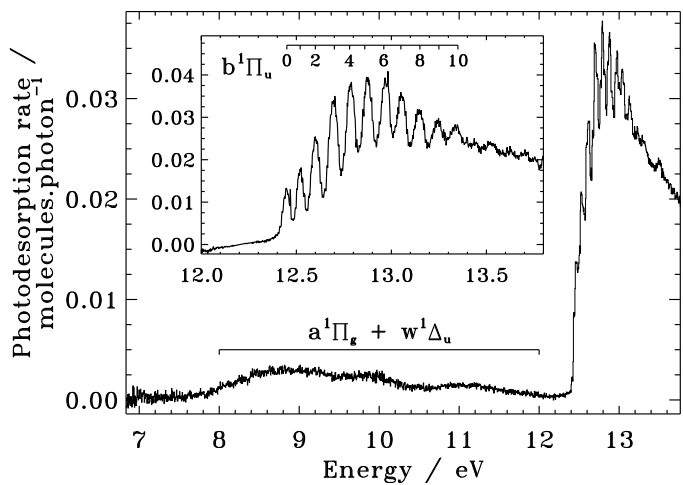}
 			\caption{Photodesorption spectrum of 60~ML-thick $\rm^{15}$N$_2$ ice at 15~K between 7 and 13.6 eV. Spectroscopic assignments from \cite{Haensel:1971bm}, adapted from $\rm^{14}N_2$ to $\rm^{15}N_2$, are overplotted.}   \label{pd_N2}
		 \end{figure}

		The photodesorption spectrum of a 60~ML-thick $^{15}$N$_2$ ice at 15~K is shown in Fig. \ref{pd_N2}. It shows a strong wavelength dependency and presents a highly structured signal above 12.4~eV, as well as a clear efficiency break of more than an order of magnitude compared to the minor shallow bands between 8 and 12~eV. The photodesorption maximum is found at 12.8~eV and reaches $3.77\pm1.13\times10^{-2}$ molecules photodesorbed per incoming photon. For energies below 12.4~eV, the photodesorption rates do not exceed $4 \times 10^{-3}$~molecules~photon$^{-1}$. At Lyman-alpha, the rate of $1.5 \pm 0.9 \times 10^{-3}$~molecules~photon$^{-1}$ found here is higher than the one derived by \cite{Oberg:2009kca} but is in agreement within the experimental uncertainties. To understand the origin of the N$_2$ photodesorption spectrum, one needs to compare it to the corresponding photoabsorption spectrum \citep{Haensel:1971bm}. The structured photodesorption signal above 12.4~eV is also observed in the N$_2$ absorption spectrum. This absorption band  corresponds to the vibrational progression of the dipole-allowed b$\rm^1\Pi_u$~$\leftarrow$X$\rm^1\Sigma_g^+$ electronic transition. The similarity between the photoabsorption and photodesorption spectra indicates that N$_2$ photodesorption is triggered by the excitation of N$_2$ in the b$\rm^1\Pi_u$ state. This transition is, however, predissociative in the gas phase due to a valence-Rydberg coupling \citep{James:1990go} and leads to the formation of atomic N($^2$D) and N($^4$S). The dissociation energy of N$_2$ in the gas phase is 9.6 eV, lower than the observed photodesorption band. However, neither $^{15}$N atoms nor $^{15}$N$_3$ molecules are detected by the QMS during ice irradiation, which suggests a photostability of N$_2$ in the ice up to 13.6~eV.\

		\begin{figure}
			\centering
 			\includegraphics[width=1\linewidth]{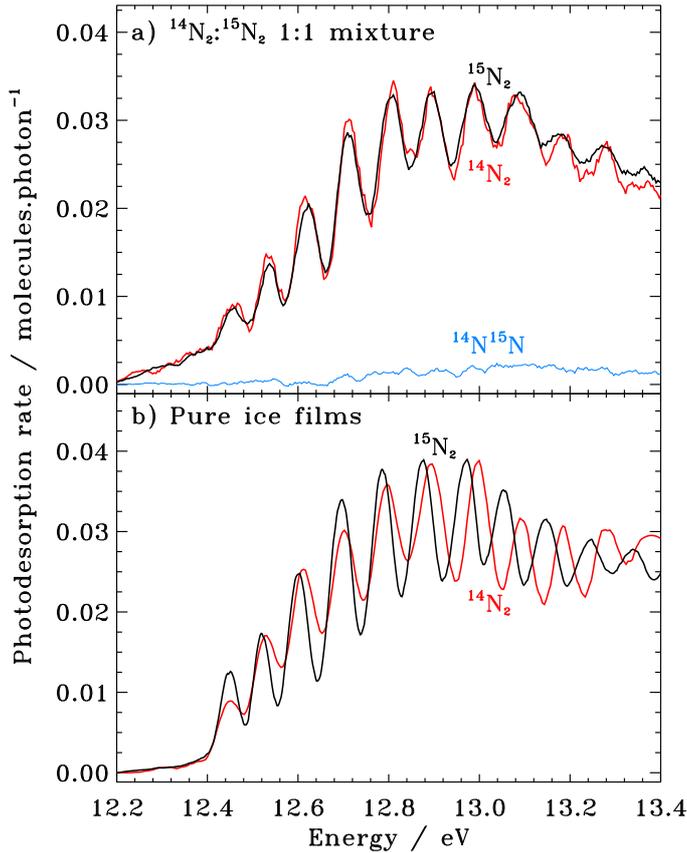}
 			\caption{Photodesorption spectra of $^{15}$N$_2$ (back line), $^{14}$N$_2$ (red line), and $^{15}$N:$^{14}$N (blue line) between 12.2 and 13.4 eV from a) a 60~ML-thick $^{15}$N$_2$:$^{14}$N$_2$ 1:1 ice at 14~K and b) pure 30~ML-thick $^{15}$N$_2$ and $^{14}$N$_2$ ice films at 14~K. The photodesorption spectra have been smoothed to facilitate comparisons between the vibrational progressions.}
			
			\label{psd_15N14N}
		 \end{figure}
		
		 To further test whether N$_2$  photodesorption could be initiated by photodissociation, a 1:1 $^{15}$N$_2$:$^{14}$N$_2$ mixture of 60~ML at 15~K was irradiated, and the photodesorption measurements of $^{15}$N$_2$, $^{14}$N$_2$, and $^{15}$N$^{14}$N are presented in Fig. \ref{psd_15N14N}a. The desorption of $^{15}$N$^{14}$N is negligible during irradiation and the low percentage detected during a TPD experiment following irradiation is consistent with the gas impurity level indicated by the $^{15}$N$_2$ gas provider. This implies an inefficient dissociation of N$_2$ or a very efficient and immediate recombination of dissociated N$_2$ due to cage effects in the solid phase. The formation of azide (N$_3$) after UV photolysis of solid N$_2$ has been reported by \cite{Wu:2012hr} in a similar UV irradiation experiment but using a higher photon flux. In the present case, additional high flux irradiation experiments with a photon distribution peaking at 13 eV (1 eV width and 2.9 $\times$ 10$^{13}$ photons~s$^{-1}$~cm$^{-2}$ intensity) were performed on a 60~ML $^{15}$N$_2$:$^{14}$N$_2$ 1:1 film and monitored using RAIRS for 40 minutes, followed by a TPD experiment. None of the N$_3$ isotopologs have been detected in the infrared spectra. This does not contradict the results from \cite{Wu:2012hr} since the fluence they employed was about 19 times higher than what is used in the present experiments and our non-detection also agrees with the non-detection reported in a H$_2$-discharge lamp-based UV irradiation study by \cite{2002ApJ...568.1095H}.\

		 Figure \ref{psd_15N14N}b presents the photodesorption spectra of $^{15}$N$_2$ and $^{14}$N$_2$ from pure 30~ML-thick ice films at 14~K. A comparison of the two spectra clearly shows an isotopic shift in the vibrational progression for both species upon excitation towards the b$\rm^1\Pi_u$ state. This supports a mechanism similar to what has been discussed for CO by \cite{Bertin:2012wi}, where similar behavior was observed for $^{12}$CO and $^{13}$CO. This isotopic shift is not visible in the 1:1 $^{15}$N$_2$:$^{14}$N$_2$ photodesorption spectra (Figure \ref{psd_15N14N}a), since here the photodesorption pattern of both isotopologs overlaps. This overlap is easily explained by the fact that in mixed ice films, the desorption of any N$_2$ molecule can be induced by both isotopologs. 
	 
		 The evidence of vibronic excitation in the N$_2$ photodesorption spectrum and the lack of recombined $^{15}$N$^{14}$N or N$_3$ isotopologues indicate that the underlying mechanism is a desorption induced by electronic transition (DIET) process. The mechanism above 12.4 eV is similar to the mechanism found for CO photodesorption below 10 eV by \cite{Fayolle:2011eh} and \cite{Bertin:2012wi} and is discussed further in Section \ref{Sec_dis}.\
		 		  
		 The minor photodesorption signal obtained for energies lower than 12 eV may be due to the excitation of the forbidden Lyman-Birge-Hopfield a$\rm^1\Pi_g$~$\leftarrow$X$\rm^1\Sigma_g^+$ and Tanaka $\rm w^1\Delta_u$~$\leftarrow$X$\rm^1\Sigma_g^+$ transitions according to \cite{Wu:2012hr}. These very weak transitions have also been reported in absorption in the solid phase by, e.g., \cite{Brith:1965ev}, \cite{Roncin:1967iq}, and \cite{Mason:2006bp}. The two vibronically resolved progressions associated to the excitation of a$\rm^1\Pi_g$ and $\rm w^1\Delta_u$ observed in absorption spectrum of solid N$_2$ \citep{Mason:2006bp} are not seen in our photodesorption spectrum. These transitions, which are both electronically dipole forbidden in the gas phase, are weak in the solid state. It is possible that these narrow bands are buried in the continuum that is visible between 8 and 12~eV in Fig. \ref{pd_N2}.

	\subsection{Oxygen photodesorption}

\begin{figure}
			\centering
 			\includegraphics[width=1\linewidth]{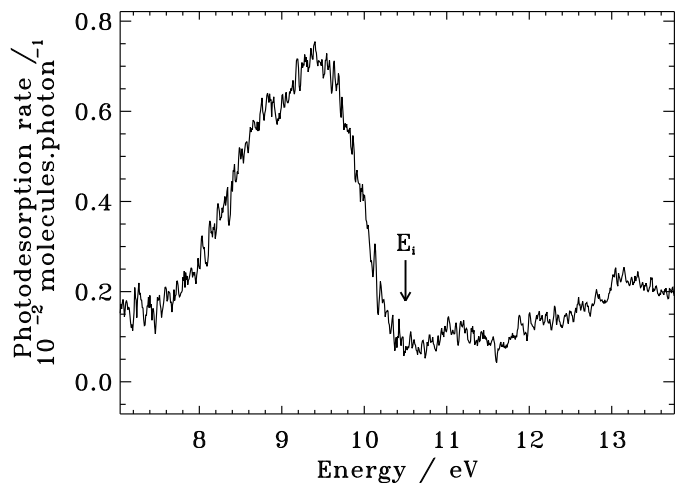}
 			\caption{Photodesorption spectrum of O$_2$ for a 30~ML ice film at 15~K between 7 and 13.6 eV. The E$_i$ arrow presents the ionization threshold O$_2$ in the solid state.}\label{psd_O2}
		 \end{figure}

	Figure \ref{psd_O2} shows the photodesorption spectrum of a 30~ML-thick O$_2$ ice at 15~K. The spectrum displays a strong, broad and unstructured desorption pattern with clearly no evidence of any vibrational structure. The photodesorption value exceeds 5~$\times$~10$^{-4}$ molecules~photon$^{-1}$ for all photon energies in the studied spectral window between 7.5 and 13.6 eV, with a starting value of $\sim$1.5~$\times$~10$^{-3}$ molecules~photon$^{-1}$ around 7.5 eV that hints for a photodesorption onset at lower energies. Molecular oxygen dissociates in the solid phase as is shown by the detection of atomic oxygen in the gas phase during irradiation. The signal of m/z=16 has been recorded during irradiation and an upper limit of the amount of desorbing atomic oxygen is derived by also taking the fragmentation pattern of O$_2$ into O$^+$ and O$_2^+$ in the QMS into account; oxygen atoms desorbing from the surface are detected with a concentration lower than 8 \% with respect to desorbing molecular O$_2$. The detection of O-atoms stemming from the UV photolysis hints at an excitation mechanism followed by dissociation. The O$_2$ and O recombination should yield ozone, but no significant m/z=48 signal is observed during irradiation or subsequent TPD. Ozone likely forms but in very small amounts considering the low fluence employed in the present experiment; the flux amounts to a few 10$^{12}$ photons~s$^{-1}$~cm$^{-2}$ during 20 minutes scanning the 7 to 13.6~eV range.\

There is a good agreement between the O$_2$ photodesorption spectrum presented here and the VUV absorption spectrum recorded by \cite{Mason:2006bp} over the 7 - 10.5~eV range, and both spectra are dominated by a broad structure peaking at 9.3 eV (see also Fig. \ref{psd_O2}). This band may link to the $\rm B^3\Sigma_u^- \, \leftarrow X^3\Sigma_g^-$ Schumann-Runge continuum as proposed by \cite{Mason:2006bp}, who discuss the difference between solid state and gas phase properties. Indeed, direct dissociation via the Schumann-Runge continuum produces O($^3$P) and O($^1$D) above 7.01~eV in the gas phase \citep{Parker:2000tk}, which is consistent with the observation of atomic oxygen desorbing from the O$_2$ ice film upon VUV irradiation. An alternative or possible parallel explanation is that the broad 9.3~eV desorption peak is associated with an excitation towards the $\rm E^3\Sigma_u^-$ transition with maximum absorption energies in the gas phase around 9.5 and 11~eV. This electronic transition would also lead to the formation of O($^1$D) atoms through dissocation \citep{Lee:2000fq}. In \cite{Mason:2006bp}, oxygen dimers (O$_2$)$_2$ have also been proposed as a possible starting point for desorption, particularly for lower energies, but this cannot be verified here.\

For energies above 10.5~eV, the molecular origin of the O$_2$ photodesorption mechanism is more ambiguous because UV absorption data are not available in the condensed phase. We observe a minimum around 10.5~eV that coincides with the ionization threshold of molecular oxygen in the condensed phase \citep{Himpsel:1975ex}. This would mean that ionic species are involved in the desorption mechanism. However, since the first ionic states are non-dissociative \citep{Yang:1994hb}, reactions involving the O$^+$ ion can be ruled out. Moreover, a fast neutralization is expected upon single ionization \citep{Avouris:1989wr}, consequently ionic states may relax to neutral states (below 10.5~eV), allowing a desorption mechanism that is fairly similar to the one observed in the neutral case. It will be interesting to perform additional experiments to confirm this scenario.	
	
\begin{figure}
			\centering
 			\includegraphics[width=1\linewidth]{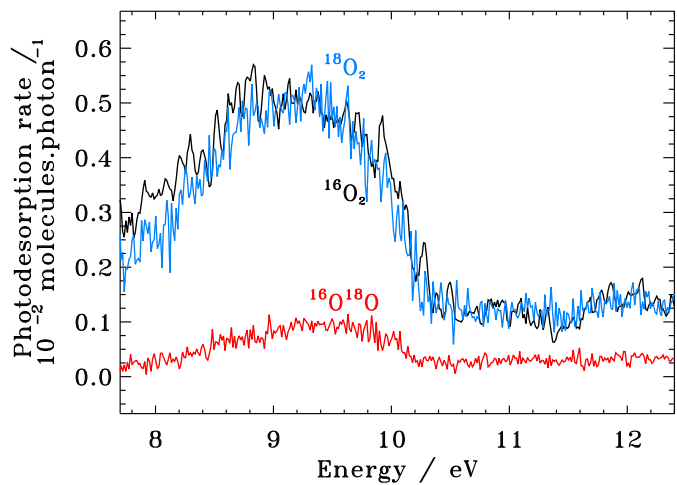}
 			\caption{Photodesorption spectrum of $^{16}$O$_2$:$^{18}$O$_2$ 1:1 at 15~K, 60~ML between 7.5 and 13.6 eV.}\label{psd_O2_diss}
		 \end{figure}

	Figure \ref{psd_O2_diss} presents the photodesorption signal of $^{16}$O$_2$, $^{18}$O$_2$, and $^{16}$O$^{18}$O from a 1:1 $^{16}$O$_2$:$^{18}$O$_2$ ice mixture at 15~K and 60~ML thick. The photodesorption signal of the initial $^{16}$O$_2$ and $^{18}$O$_2$ isotopologs overlap and a contribution of the $^{16}$O$^{18}$O isotopolog, stemming from the dissociation and recombination of the initial mixture components, is clearly visible. This shows that the O$_2$ photodesorption mechanism involves non-negligible channels due to O$_2$ dissociation.\

\begin{figure}
			\centering
 			\includegraphics[width=1\linewidth]{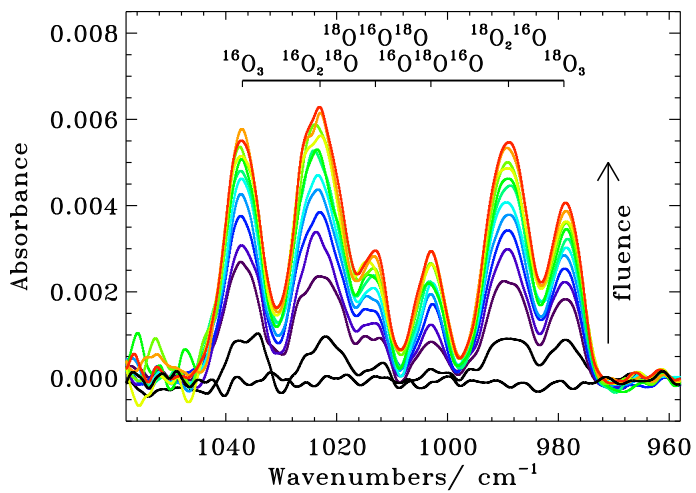}
 			\caption{RAIRS spectra of O$_3$ isotopolog formation for increasing UV fluence (from black to red) from a 60~ML thick $^{16}$O$_2$:$^{18}$O$_2$ 1:1 ice irradiation experiment at 9.2 eV.}\label{spec_O3}
		 \end{figure}

	When irradiating the 60~ML thick 1:1 $^{16}$O$_2$:$^{18}$O$_2$ mixture at 15~K with 9.2 eV photons using the undispersed radiation (1 eV FWHM) mode with high photon flux (4.3 $\times$ 10$^{14}$ photons~s$^{-1}$~cm$^{-2}$ at 9.2 eV), the formation of ozone is observed by RAIRS. Figure \ref{spec_O3} shows all O$_3$ isotopologs forming during the irradiation of an 1:1 $^{16}$O$_2$:$^{18}$O$_2$ ice mixture. The assignment of the various isotopologs is possible by using gas phase data obtained by \cite{Dimitrov:1998gj}. As shown by the presence of the six ozone bands, all combinations of $^{16}$O and $^{18}$O are formed. O atoms attack molecular oxygen to produce significant amounts of O$_3$, and the formation of $^{16}$O$^{18}$O$^{16}$O and $^{18}$O$^{16}$O$^{18}$O can result either from an addition or inclusion of atomic oxygen to molecular oxygen. Gas phase experiments show a strong isotopic dependence of the formation of ozone \citep{Janssen:1999id,Mauersberger:2003wm}, and recent experiments on ozone synthesis in the solid state induced by electron bombardment yield an even stronger isotopic enrichment \citep{Mebel:2010vg} in favor of the heavier isotopolog. The O$_2$ photodesorption rate will depend directly on the branching ratio between desorption and dissociation, so more detailed studies on this isotope effect are warranted.\
	
\section{Discussion} \label{Sec_dis}

\subsection{Photodesorption mechanisms for pure ices}

Photodesorption is a wavelength-dependent process. The similarities between the absorption and the photodesorption spectra for N$_2$, O$_2$, and CO \citep{Fayolle:2011eh,Bertin:2012wi} clearly show that photodesorption in the VUV is induced by electronic excitation of the condensed molecules. The nature of the excited state reached after UV irradiation determines the mechanism leading to photodesorption. Two different families of molecules can be distinguished: a first class where molecules do not dissociate (e.g., CO, N$_2$) and a second class where dissociation occurs (e.g., O$_2$, H$_2$O, CH$_3$OH, e.g., \citealt{1995Natur.373..405W,Oberg:2009cd}). The photodesorption of both types is triggered by the initial photon absorption, but they differ in the energy redistribution mechanisms from excited to desorbing species. Desorption induced by ionization is not discussed here since the ionization potential of most simple molecules found in interstellar ices is above or at the high end of the investigated energy range.\

In the case where absorption of a UV photon does not lead to dissociation, excited molecules can induce a sudden rearrangement within the ice lattice when returning to the ground state and eject some of the surface molecules into the gas phase. This indirect DIET process has been highlighted in \cite{Bertin:2012wi}, where a layer of $^{13}$CO deposited on a thicker $^{12}$CO ice film was irradiated between 7 and 13.6 eV. Molecules desorbing in this case are mainly $^{13}$CO, but they follow the excitation pattern of subsurface $^{12}$CO. The same mechanism is observed in the case of N$_2$ for the excitation to the b$\rm^1\Pi_u$ state. The photodesorption efficiency of this subsurface excitation inducing surface desorption mechanism is thus directly linked to the efficiency of the energy redistribution between the excited molecule and the surface. Investigating the energy-coupling efficiency between different kinds of molecules is crucial since the excitation of one species could induce the desorption of another.\

In the second case where molecules are dissociated after UV photon absorption, there are additional pathways to bring condensed molecules into the gas phase. Similar to what has been seen in molecular dynamic simulations for the case of H$_2$O  \citep{Andersson:2008bo,Andersson:2011ce}, the formation of oxygen radicals affects the O$_2$ photodesorption mechanism. O$_2$ photodesorption can come from

\begin{easylist}
\ListProperties(Numbers1=a,Hide2=1,Numbers2=r,Numbers3=a,Hide3=2,Progressive*=.5cm, FinalMark3={)}, FinalMark2={)}, FinalSpace=.3em)

&  Photon absorption \ce{O_{2(s)} ->[\ce{h\nu}] O^*_{2(s)}};

&  Redistribution of the energy that induces

&& an indirect DIET process \ce{O^*_{2(s)} ->  O_{2(g)}};

&& O$_2$ dissociation  \ce{O^*_{2(s)} ->  O + O}

&&& \ce{O + O -> O_{2(g)}} \\exothermic recombination;

&&& \ce{O_{2(s)} + O -> O_{2(g)} + O} \\kick-out by atomic oxygen;

&&& \ce{O_{2(s)} + O -> O_3_{(s)} ->[\ce{h\nu}] O_{2(g)} + O} \\ photoinduced dissociation of ozone.
\end{easylist}
\

The additional photodesorption pathways accessible by the formation of atomic oxygen are ii) 1) recombination of oxygen radicals to reform molecular oxygen followed by desorption to dissipate excess of energy, ii) 2) kick out of molecular oxygen from the surface induced by mobile atomic oxygen, and ii) 3) reaction of atomic oxygen with a molecular oxygen to form ozone, which can be photolyzed into O$_2$ with enough energy to desorb. The recombination of oxygen atoms, reaction ii) 1), is related to the probability that two oxygen atoms meet and react. This reaction channel depends highly on the diffusion of oxygen atoms, as well as on their concentration in the ice, thus it is expected to be flux dependent. In addition, the photodesorption channel involving the photolyzis of ozone, ii) 3), will only be effective for a high photon dose. In the low flux experiments presented here, very little O$_3$ is formed compared to the high flux experiments. Additional studies investigating O$_2$ photolyzis and desorption with respect to the photon flux are required to confirmed and quantify a flux dependency of each reaction channels.

\subsection{Astrophysical consequences}

\begin{table*}[ht]

\caption{Photodesorption rates of pure N$_2$, O$_2$, and CO ice for different ISM environments.}             
\label{tab_yields} 
\begin{minipage}[c]{\linewidth}
\centering
\begin{tabular}{l c c c }       
\hline\hline  
Environment 	& N$_2$ & O$_2$ & CO\\
			&  molecules~photon$^{-1}$ & molecules~photon$^{-1}$ & molecules~photon$^{-1}$\\
\hline			
Edges of clouds \footnote{using the ISRF from \cite{Mathis:1983vl}} & 2.6 $\times$ 10$^{-3}$ & 3.3 $\times$ 10$^{-3}$  & 1.3 $\times$ 10$^{-2}$\\
Prestellar cores\footnote{using the spectrum calculated by \cite{Gredel:1987vz}} & 2.2 $\times$ 10$^{-3}$ &  2.6 $\times$ 10$^{-3}$ & 1.0 $\times$ 10$^{-2}$\\
Black body 10,000 K& 5.3 $\times$ 10$^{-3}$ & 2.3 $\times$ 10$^{-3}$ &  6.4 $\times$ 10$^{-3}$ \\
TW Hydrae \footnote{using the spectrum collected by \cite{Herczeg:2002es}, \cite{Valenti:2003dl}, and \cite{JohnsKrull:2007eq}} &2.1 $\times$ 10$^{-3}$ & 2.3 $\times$ 10$^{-3}$ & 7.2 $\times$ 10$^{-3}$ \\
Lyman-alpha &1.5 $\times$ 10$^{-3}$& 2.1 $\times$ 10$^{-3}$ & 4.2 $\times$ 10$^{-3}$\\

\hline
\end{tabular}
\end{minipage}
\end{table*}

The measured photodesorption spectra of N$_2$, O$_2$, and CO \citep[present study and][]{Fayolle:2011eh,Bertin:2012wi} is wavelength dependent, which implies that the photodesorption efficiency in space will depend on the FUV field profile. Photodesorption in the ISM is expected to be the most efficient in the 7.0~--~13.6~eV range. The 7.0~eV lower limit of this range corresponds to the minimum energy required for an electronic transition of most small molecules found in the ISM, while the 13.6 eV upper limit corresponds to the Lyman edge where radiation in molecular clouds is absorbed by hydrogen.
Table \ref{tab_yields} exemplifies how the photodesorption rates of N$_2$, O$_2$, and CO can change with respect to the FUV field encountered in different environments. Values are obtained by convoluting the FUV profiles to the photodesorption spectra between the 92 and 172 nm (7.2 - 13.6 eV). In other words, these calculations weight the photodesorption rate derived from the laboratory experiments by the fraction of photons at discrete energies in various ISM regions. For the calculations performed in Table \ref{tab_yields}, the FUV profile for the interstellar radiation field at the edge of the molecular clouds is provided by \cite{Mathis:1983vl}, the Lyman and Werner series used for the prestellar cores calculation are taken from \cite{Gredel:1987vz}, and the TW hydrae spectrum used here has been collected by \cite{Herczeg:2002es}, \cite{Valenti:2003dl}, and \cite{JohnsKrull:2007eq}. A 10000~K black body curve is taken to approximate the emission from a Herbig Ae star, and photodesorption rates at pure Lyman-alpha are presented as well. Values presented for CO are similar to those reported by \cite{Fayolle:2011eh} except for the 10000~K black body case where the Planck equation used in the previous study was incorrect.\

In general, the photodesorption rates calculated here are significantly higher than those employed in current astrochemical models, especially for CO \citep[e.g.][]{2011A&A...534A..49G,2013ApJ...762...86V}. This is the case because most experimental values derived in the past were obtained using an hydrogen discharged lamp emitting photons mainly at Lyman-alpha. As proven by the photodesorption spectra obtained experimentally, it appears that photodesorption rates are in general lower at this particular wavelength than at other FUV wavelengths.\

When comparing CO, N$_2$, and O$_2$, the change in photodesorption rates between the different ISM environments is most significant for CO ice between the edge of the molecular clouds and when exposed to radiation at pure Lyman-alpha. In the former case, many of the UV photons are in resonance with CO A$^1\Pi$ electronic state, while CO photodesorption has a minimum at 10.2~eV. N$_2$ and O$_2$ photodesorption rates vary less in the FUV (less than a factor 4) and have a lower rate than CO in all regions. This implies that in regions where the main desorption mechanism is photodesorption, the gas-to-pure ice ratio will be higher for CO than for N$_2$ or O$_2$.\

Care has to be taken when using the present laboratory spectra to derive photodesorption efficiencies with respect to the local FUV field since these values are only given for pure ices. Where ices are mixed, detailed chemical modeling that takes the excitation-desorption mechanism of each species into account has to be used to predict photodesorption efficiencies. For species that do not dissociate (e.g., N$_2$, CO), the subsurface molecules will induce the desorption of surface molecules, while in the case of dissociation (O$_2$, H$_2$O), the radicals created can induce chemistry that will affect photodesorption efficiencies. For example, \citep{Hollenbach:2008ho} use a value of 10$^{-3}$~molecules~photon$^{-1}$ for the photodesorption rate of O$_2$, which is very close to the value derived experimentally for pure O$_2$ ice. In mixed ices, however, the photodesorption of O$_2$ could be perturbed: on one hand, the atomic oxygen produced upon irradiation can recombine with other molecules or radicals, lowering photodesorption induced by exothermic oxygen recombination; on the other hand, the presence of additional excited species and radicals may enhance indirect O$_2$ photodesorption via DIET or kick-out mechanisms. It is not obvious how these effects could balance each other. More experiments and molecular dynamics simulations are required to treat photodesorption in mixed ices accurately. The present study offers a firm first step in this direction.

\section{Conclusions}

The FUV induced photodesorption of pure N$_2$ and O$_2$ ices are principally different. N$_2$ does not significantly dissociate upon irradiation between 7 and 13.6 eV. Its transition to the first allowed electronic state (above 12.4 eV) induces desorption through an indirect process where the excited molecules eject surface molecules into the gas phase. In the case of O$_2$, photodissociation already occurs below 7 eV, and oxygen radicals induce chemical pathways driving the photodesorption mechanism. The formation of ozone in the case of high flux irradiation affects the photodesorption of O$_2$ by adding an additional channel for molecular oxygen desorption induced by ozone photolyzis. Absolute wavelength-dependent photodesorption rates can be used to estimate photodesorption in various star-forming environments, but it is important to note that the presented results apply to pure ices only. A detailed modeling of neighbor interaction and radical-induced chemistry is required to predict photodesorption efficiencies in the case of mixed ices.

\begin{acknowledgements}
We are grateful to Ewine van Dishoeck and Marc van Hemert
for stimulating discussions and to the anonymous referee for helpful comments and suggestions. We thank Nelson De Oliveira for assistance on the beamline and to SOLEIL for providing beamtime under project 20110823 and to the general SOLEIL staff for smoothly running the facility. Financial support from the French national
program PCMI (Physique chimie du milieu interstellaire), the UPMC platform for astrophysics "ASTROLAB", the
Dutch program NOVA (Nederlandse Onderzoekschool Voor Astronomie),
and the Hubert Curien Partnership Van Gogh are gratefully
acknowledged.
\end{acknowledgements}

\bibliographystyle{aa}
\bibliography{mybib}

\end{document}